\documentclass[onecolumn, 8pt]{webofc}
\usepackage[varg]{txfonts}   % Web of Conferences font
\usepackage{algorithm}

\usepackage[ddmmyyyy]{datetime} %Displaying last compiled time 
\usepackage{amsmath} % Package for \eqref
\usepackage[dvipsnames]{xcolor}% Package for \textcolor
\usepackage{cancel} % Package for \cancelto{0}{x}
\usepackage{graphicx}
\usepackage{comment} %SA added 14/01/2019
\usepackage{soul,ulem} %For striking out - PK: March 28, 2019 
\soulregister\cite7
\soulregister\ref7 
\soulregister\eqref7 %To abose soul to make \hl through cite, ref, and eqref % https://tex.stackexchange.com/questions/139463/how-to-make-hl-highlighting-to-automatically-place-incompatible-commands-in
\usepackage{bbding}
\usepackage{pifont}
\usepackage{wasysym}
\usepackage{amssymb}
\usepackage{textcomp} % added by PK on Oct 9 2019 for the  \textrightarrow \textdownarrow 
\usepackage[colorinlistoftodos
, textsize=tiny]{todonotes} %\todo{text}
\usepackage{textcomp} % added by PK on Oct 9 2019 for the  \textrightarrow \textdownarrow 
\usepackage{lineno} % added by PK on Oct 18 for the line number
\usepackage{hyperref}
\usepackage{cleveref}
\hypersetup{
    colorlinks=true,
    citecolor=[rgb]{0.85,0.17,0.17},
    linkcolor=blue,
    filecolor=magenta,      
    urlcolor=cyan,
    urlbordercolor=red
}

\newcommand{\gNtilde}{\ensuremath{\tilde{\gamma}_{\rm N}}}

\newcommand{\GN}{\ensuremath{G_{\rm N}}}
\newcommand{\M}{\ensuremath{M_{500}}}
\newcommand{\R}{\ensuremath{R_{500}}}

\newcommand{\ksM}{\text{km/s Mpc$^{-1} $}}
\newcommand{\Psz}{\ensuremath{P_{\rm SZ}}}
\newcommand{\Px}{\ensuremath{P_{\rm X}}}
\newcommand{\Tx}{\ensuremath{T_{\rm X}}}

\defcitealias{Ettori18}{E19}

\definecolor{mypurple}{RGB}{135,10,190} % purple {135,10,110}

\begin{document}
\title{Generalised scalar-tensor theories of gravity and pressure profiles of galaxy clusters}
%
% subtitle is optional

\author{
\firstname{Balakrishna S.} \lastname{Haridasu}\inst{1,2,3}\fnsep\thanks{\email{sandeep.haridasu@sissa.it}} 
    \and 
	\firstname{Purnendu} \lastname{Karmakar}\inst{4} 
	\and
	\firstname{Marco} \lastname{De Petris}\inst{4,5,6}
	\and
	\firstname{Vincenzo F. } \lastname{Cardone}\inst{5,6}
	\and
	\firstname{Roberto } \lastname{Maoli}\inst{4,5}
        % etc.
}

\institute{SISSA-International School for Advanced Studies, Via Bonomea 265, 34136 Trieste, Italy
\and
          INFN, Sezione di Trieste, Via Valerio 2, I-34127 Trieste, Italy
\and
          IFPU, Institute for Fundamental Physics of the Universe, via Beirut 2, 34151 Trieste, Italy
\and
          Dipartimento di Fisica, Sapienza Universit\'a di Roma, P.le Aldo Moro 2, 00185, Roma, Italy
\and
          INAF - Osservatorio Astronomico di Roma, Via Frascati 33, 00040, Monteporzio Catone, Roma, Italy    
\and
            INFN- Sezione di Roma 1, P.le Aldo Moro 2, 00185, Roma, Italy
% etc.
          }

\abstract{%
  In the current proceedings, we summarise the results presented during the mm Universe@NIKA2 conference, taken from our main results in \cite{Haridasu:2021hzq}. We test the Degenerate higher-order scalar-tensor(DHOST) theory as a generalised platform for scalar-tensor theory at galaxy cluster scales to predict in such static systems small scale modification to the gravitational potential. {DHOST theory is not only a good alternative to $\Lambda$CDM for the background evolution but also predicts small-scale modification to the gravitational potential in static systems such as galaxy clusters.} With a sample of 12 clusters with accurate X-ray Intra Cluster Medium (ICM) data (X-COP project) and Sunyaev-Zel'dovich (SZ) ICM pressure (Planck satellite), we place preliminary constraints on the DHOST parameters defining the deviation from GR. Moreover, we also collect a few supplementary analyses we have performed during the course: i) Gaussian process reconstruction without parametric assumptions, ii) $\Psz$-only data analysis not aided by the X-ray data. Finally, we present possible extensions to the current work which may benefit from future high sensitivity and spatial resolution observations such as the ongoing NIKA2 camera.
}
\date{\today}

\maketitle
\section{Introduction}
% \label{sec:intro}
% \section{Method and Data}
\label{sec:data} 
Einstein developed the standard general relativity (GR) by considering gravity as warps and curves in the fabric of geometric space-time, and that geometry of space-time is  described only by a metric. 
GR can explain the majority of the cosmological observations, but challenged to explain the acceleration in the cosmological distance. Therefore, if we plan to keep the basic metric structure because of the large success of GR, and add one scalar degrees of freedom to explain the acceleration, then the degenerate higher-order scalar-tensor (DHOST) theory is the best possible platform to test the largest classes of scalar-tensor theories \cite{Crisostomi:2016czh, Achour:2016rkg}. 

The additional degree of freedom must be suppressed in within the small scales to explain the observations, is the so-called screening mechanism \cite{Babichev:2013usa}. The modified Newtonian potential for the DHOST theory in the galaxy cluster, which produces the gravitational waves of the velocity ($c_g$) equal to the velocity of light ($c$), is 

\begin{equation}
\frac{d\Phi(r)}{dr} = \frac{G_{\rm N}^{\rm eff} M_{\rm HSE}(r)}{r^2} + \Xi_{1} G_{\rm N}^{\rm eff} \frac{d^2M_{\rm HSE}(r)}{dr^2}\,,
\label{eq: dphidr}
\end{equation}
while assuming the hydrostatic equilibrium (HSE), where $M_{\rm HSE}(r)$ is the total mass within the radial distance $r$. Both, $\Xi_{1}$ and $G_{\rm N}^{\rm eff}$ parameters track the departure of DHOST theory from GR, which modify the ICM thermal pressure profile as
\begin{equation}
P^{\rm th}(r) = P^{\rm th}(0) - \int_0^r \rho_{\rm gas}(\tilde r) \frac{d\Phi(\tilde r)}{dr}{\rm d} \tilde r \,,
\label{eq:pressure:profile:sz}
\end{equation}
where $\rho_{\rm gas} (r)$ is the gas density. Our formalism is a straight-up implementation of the so called \textit{forward}-method, where the pressure profile is computed while assuming empirical profiles for the mass, $M_{\rm HSE}(r)$, and the electron density, $n_e(r)$, radial profiles which in our case are the standard NFW \cite{Navarro:1995iw} and simplified Vikhlinin parametric model \cite{Vikhlinin:2005mp}, respectively. In this context, several previous works have implemented similar approach either using stacked clusters and/or having complementary weak lensing data \cite{Sakstein:2016ggl, Wilcox:2015kna, Terukina:2013eqa}.

Throughout the current discourse we assume $H_0 = 70 $ \ksM  and $\Omega_{\rm m } = 0.3$. $\R$ and $\M$ carry usual definitions, i.e, total mass $M_{\Delta}$ within the radius $R_{\Delta}$, with a mean mass density $\Delta$ times the critical density ($\rho_{\rm c}( z) =  3H^2(z)/ 8\pi \GN$). 

\section{Main Results}
\label{sec:results}

\begin{figure}[h]
\includegraphics[scale=0.35]{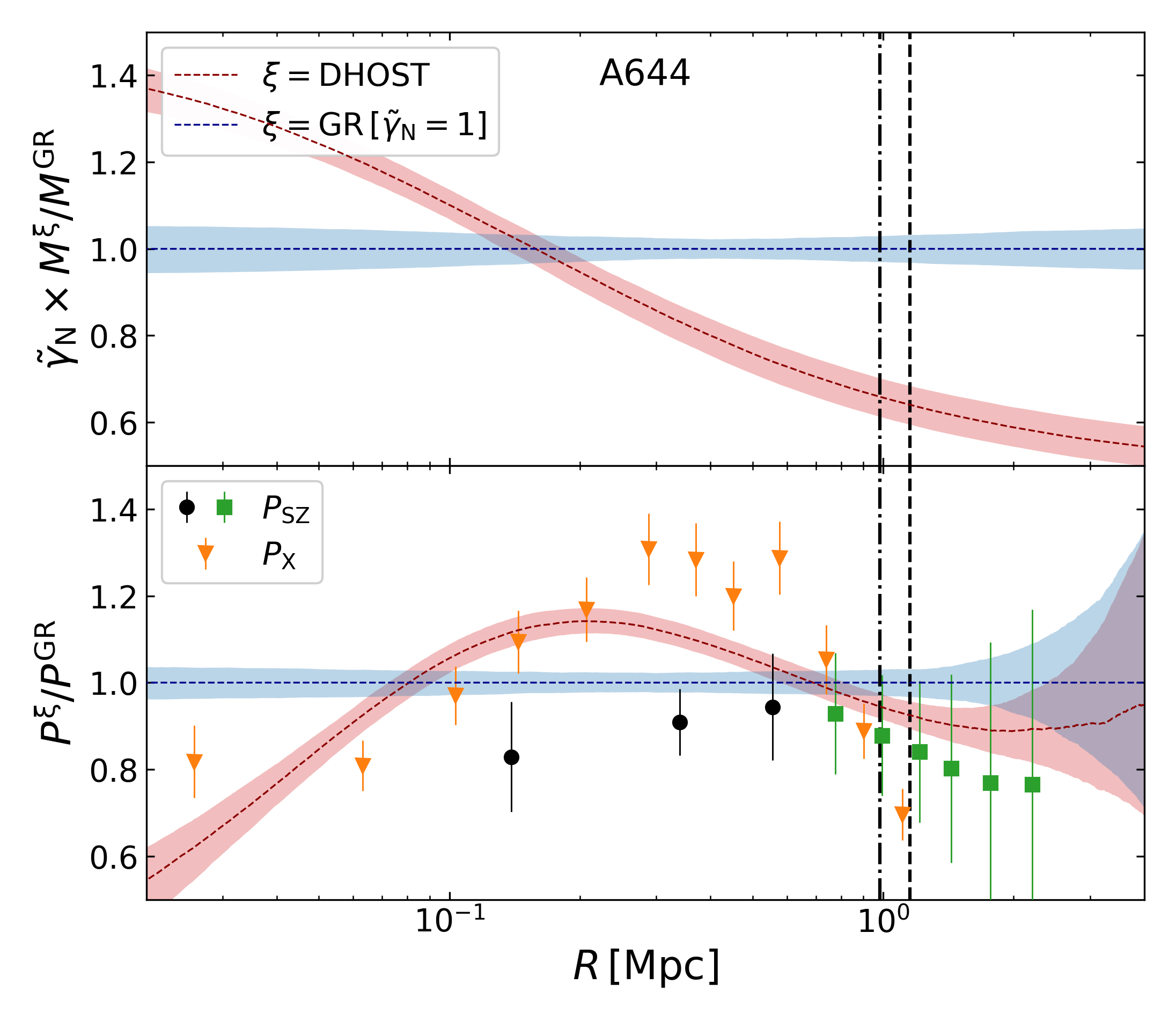}
\vspace{0.1cm}
\includegraphics[scale=0.35]{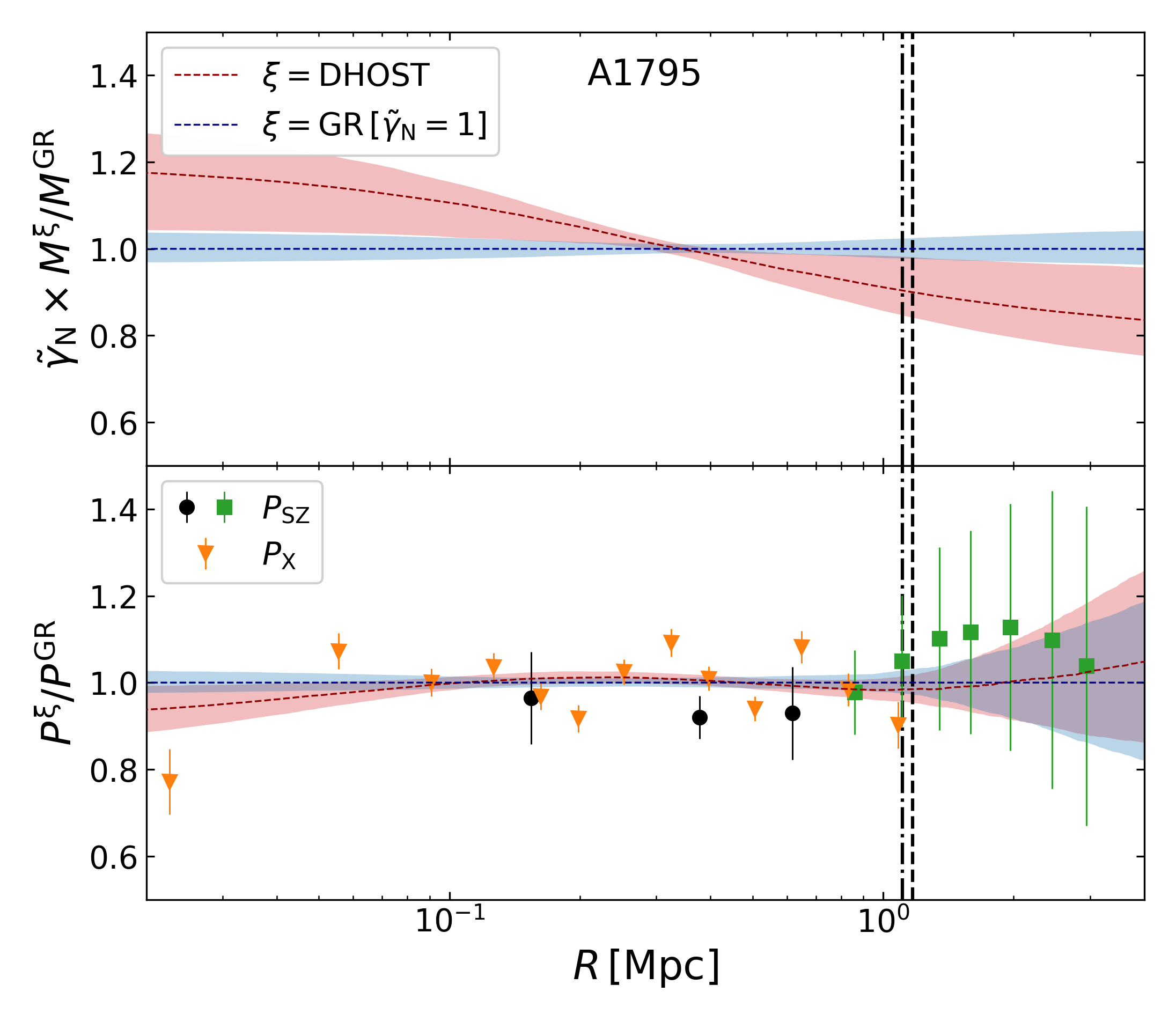}
     % Give a unique label
\caption{Mass (\textit{upper}) and pressure (\textit{lower}) radial profiles for the cluster A644 (\textit{Left}) and A1795 (\textit{Right}) in the cases of GR (blue) and DHOST (red), normalised to GR. The former cluster is an example case of the 4 non-NFW clusters, while the latter is one of the 4 clusters we retain in estimating our final conservative result. The vertical dashed (GR) and dot-dashed (DHOST) lines show $\R$. The 3 inner $\Psz$ data points (black) are excluded in the main analysis. }
\label{fig:data-fit} 
\end{figure}

\begin{figure}[h]
    \centering
    \includegraphics[scale=0.355]{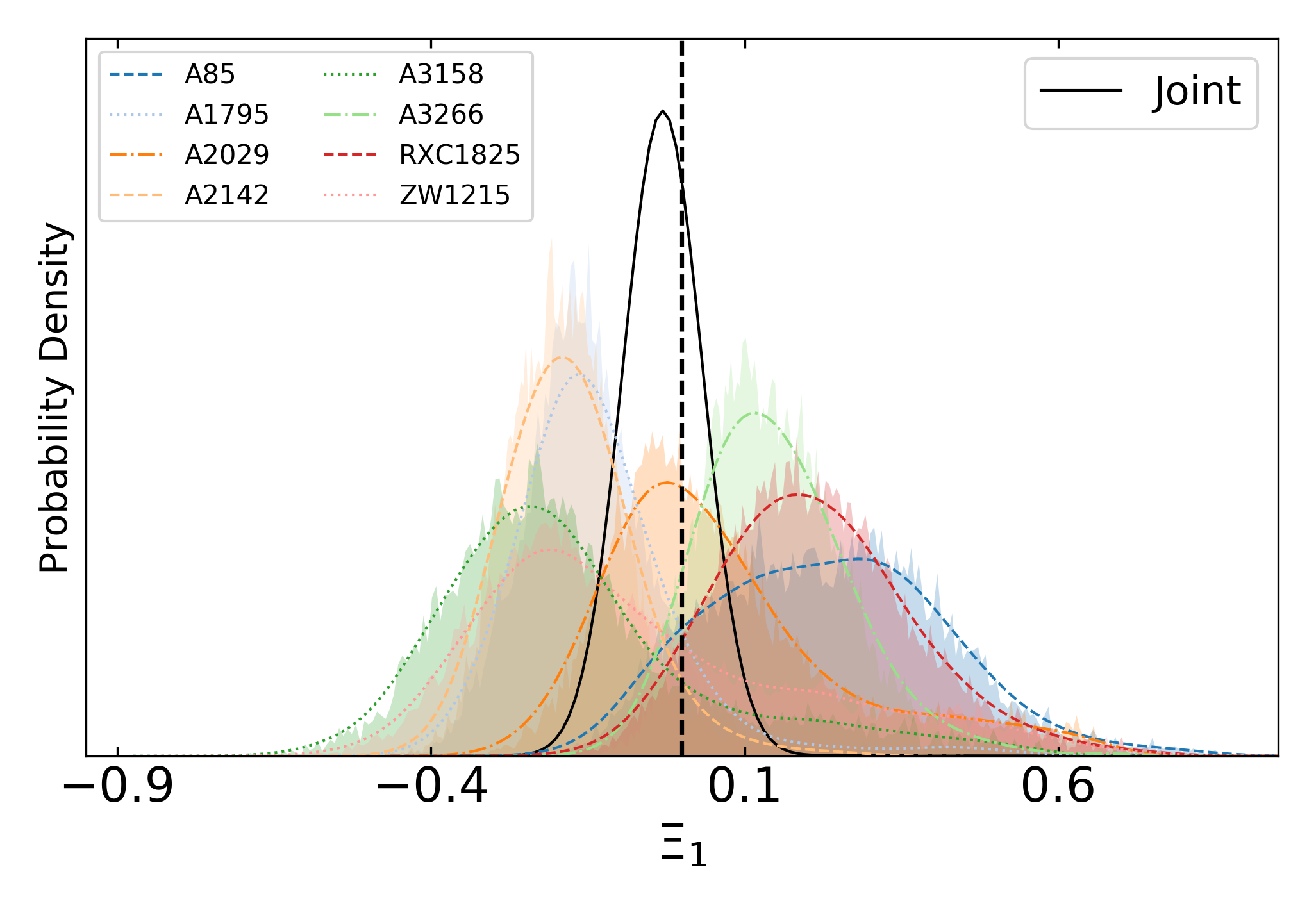}
    \vspace{-0.2cm}
    \includegraphics[scale=0.325]{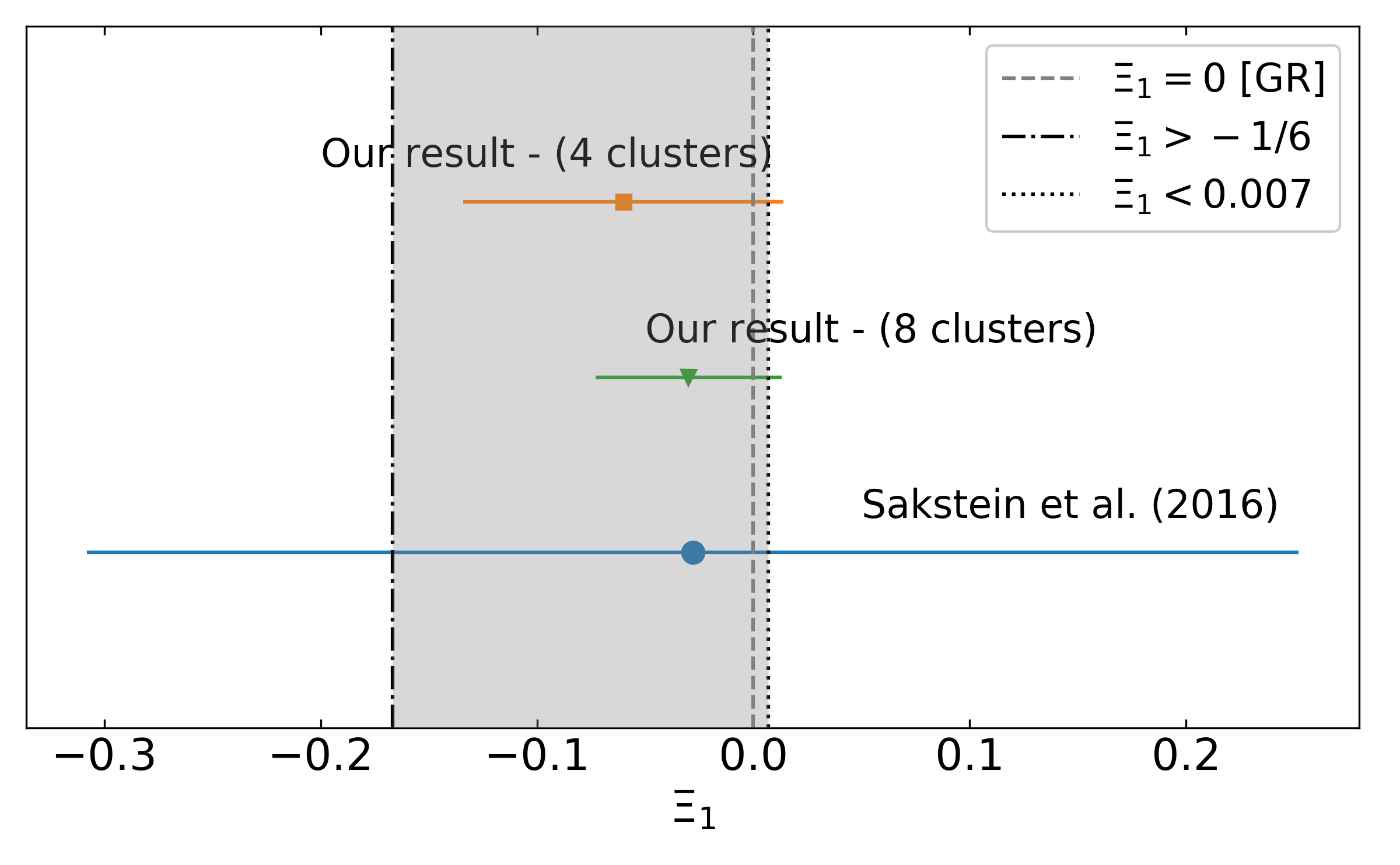}
    \caption{\textit{Left}: Posterior distribution of the parameter $\Xi_1$, over-plotted by smooth Gaussian kernel density profiles. The vertical dashed line marks the GR ($\Xi_1 = 0$) case. The black curve represents our joint constraint using 8 clusters. \textit{Right}: Constraints on the parameter $\Xi_1$, from an earlier analysis of 58 stacked clusters (blue) \citep{Sakstein:2016ggl} and our results with 8 clusters (green) and conservative 4 clusters (orange). The grey shaded region shows the allowed region for the same parameter obtained from theoretical limits \cite{Sakstein:2015aac, Sakstein:2015zoa, Saito:2015fza}.}
    \label{fig:dist}
\end{figure}

Firstly, we figure that only 8 clusters out of the 12 available clusters in the dataset prefer the NFW mass profile in a comparison through information criteria (see Fig 2, of \cite{Ettori:2018tus}). We refer to the other 4 clusters (A644, A1644, A2319 and A2255) as non-NFW clusters for brevity. As a consequence of which, we find that the non-NFW clusters fit the data within the DHOST scenario with statistical preferences reaching $\Delta\log (\mathcal{B}) >> 25$ (further details in \cite{Haridasu:2021hzq} and \cite{Ettori:2018tus}), which imply a very high preference yet providing very low values of the parameter $\Xi_1$, in complete disagreement with the theoretical limits \cite{Saito:2015fza, Sakstein:2015aac, Sakstein:2015zoa}. In \Cref{fig:data-fit}, we show the fits of the GR and DHOST scenarios for A644 (non-NFW) and A1795 cluster and the $1\sigma$ dispersion to provide a comparison. We show the posteriors for $\Xi_1$ obtained using the individual cluster and the joint constraint in left panel of \Cref{fig:dist} and a comparison with earlier constraints in the right panel. 

% \footnote{Please refer to the main paper \cite{Haridasu:2021hzq}, for details on the Bayesian evidence and the implications. }

As the main result of our analysis we updated the limits on the modified gravity parameter $\Xi_1= -0.030\pm 0.043 $, which is a much more stringent constrain in comparison to the earlier result of $\Xi_1 \sim - 0.028^{+0.23}_{-0.17}$, reported in \cite{Sakstein:2016ggl} using a similar formalism with stacked X-ray cluster profile. In a more conservative case where we utilise only 4 clusters (A1795, A3158, RXC1825 and ZW1215), we find a less stringent $\Xi_1 = -0.061 \pm 0.074$, yet twice as tighter constraint than the earlier analysis. These 4 conservative clusters are restricted within a redshift range of $0.0597<z<0.0766$. In the right panel of \Cref{fig:dist}, we show a comparison of the earlier constraints, our results and the theoretical limits. As can be seen, our results are in excellent agreement with the latter and an improvement for the constraints on $\Xi_1$. The lower theoretical limit is marked by the requirement of stable static solutions of non-relativistic stars \cite{Saito:2015fza, Saltas:2018mxc} and the upper limits are obtained on the based on the minimum mass required for hydrogen burning in low mass red dwarfs \cite{Sakstein:2015aac, Sakstein:2015zoa}.

An added advantage of our current analysis is that having individual clusters spread over a redshift range of $0.04 < z < 0.1$, we are able to present for the first time the necessary assessment of time-evolution of the parameter $\Xi_1(a\equiv 1/(1+z))$. We utilise a simple first order Taylor expansion of $\Xi_1(a)$ as the scale factor, $a\to 1$ and perform a post-MCMC analysis on the inferred posteriors of the $\Xi_1$ obtained for individual clusters. Using 8 clusters we find a mild deviation from a constant behaviour for $\Xi_1$, at present is however driven only by two clusters (A85 and A2142)and the advantage of extending the analysis over a larger dataset that covers a wider redshift range is clear. 

% \begin{figure}[h]
%     \centering
%     \includegraphics[scale=0.39]{constrain_comparison.png}
%     \caption{We show a comparison of the constraints on the parameter $\Xi_1$, form the earlier analysis of 58 stacked cluster (blue) \citep{Sakstein:2016ggl} and our results with 8 clusters (green) and conservative 4 cluster (orange). The grey shaded region shows the allowed region for the same obtained from theoretical limits \cite{Sakstein:2015aac, Sakstein:2015zoa, Saito:2015fza}. \SH{We could move this plot to the main paper, to make some space here !!} }
%     \label{fig:constraints}
% \end{figure}

\section{Supplementary Results}
In this section we present a few supplementary analyses we have performed, also constituting possible future extension of the current work.  

\begin{figure}[h]

\includegraphics[scale=0.35]{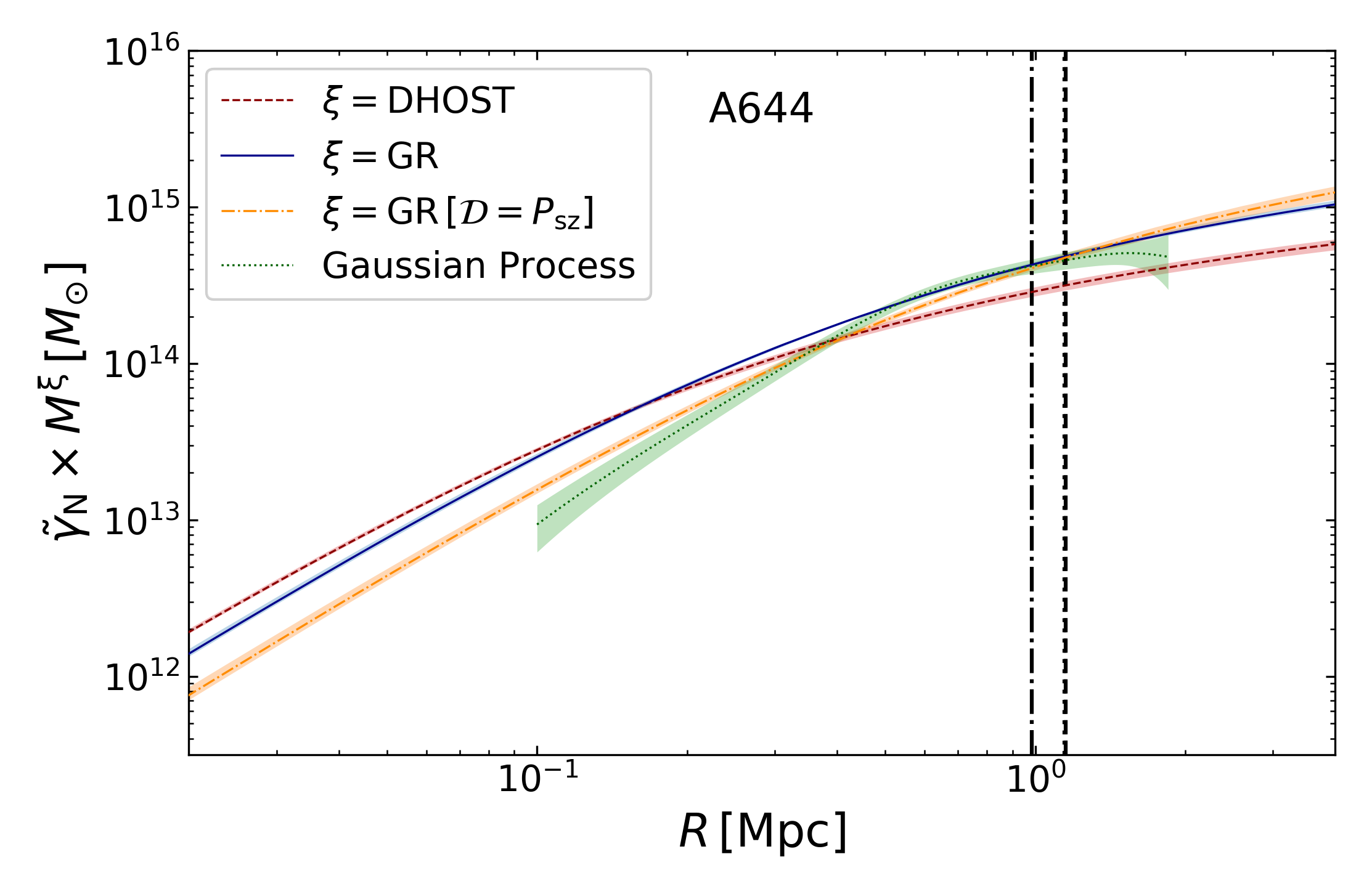}
\vspace{0.1cm}
\includegraphics[scale=0.35]{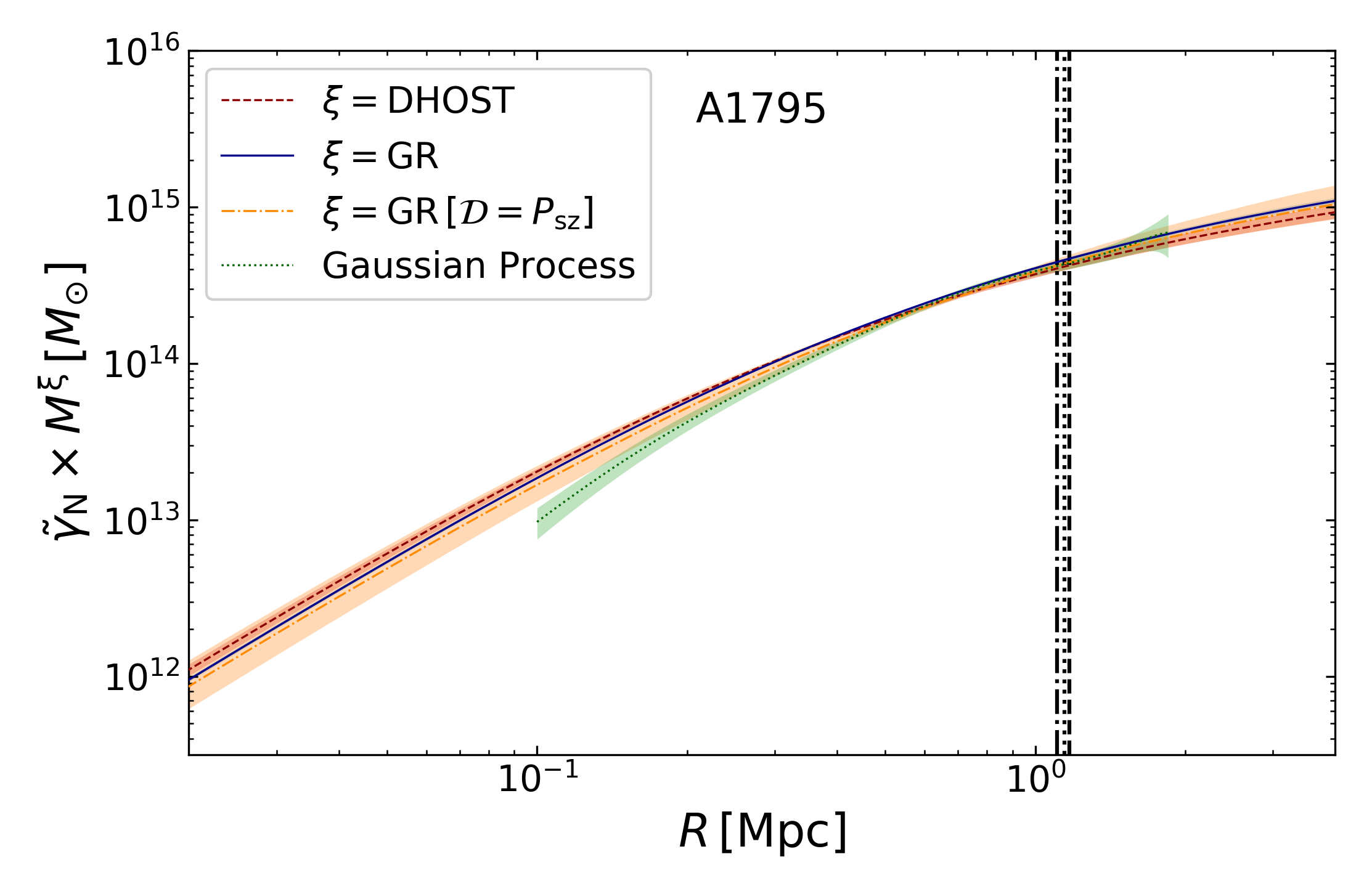}
      % Give a unique label
\caption{ Mass radial profiles with $1\sigma$ dispersion for clusters A644 (\textit{left}) and A1795 (\textit{right}) for the GR (blue), DHOST (red) and GP (green). A644 and A1795 represent one of the non-NFW cluster and one of the 4 conservative clusters, respectively. The vertical dashed, dot-dashed and dotted lines show the $\R$ in GR, DHOST and GP, respectively. The $\R$ from the GP case coincides very well with that in GR case. The mass profiles obtained utilising only the $\Psz$ data for the GR case are shown in orange. Here the GR case corresponds to $\gNtilde =1$. }
\label{fig:data-fit-GP} 
\end{figure}

\subsection{Gaussian Process }
In addition to the parametric analysis, we also perform a non-parametric model-independent analysis based on the Gaussian Process (GP) formalism, please see \cite{Haridasu:2018gqm} and references therein for details on the method. In essence, once could reconstruct the underlying functional form [$f(x_i)$] for a collection of Gaussian data points at $\mathbf{x}\equiv \{x_1, x_2, .. x_N\} $, assumed belonging to the same process, which now represents a GP. This is implemented by modelling a covariance $K(x_i,x_j)$, amongst the data points, instead of functional form for the process itself. For this purpose we independently reconstruct the electron density and the $\Psz$ pressure profiles assuming the squared-exponential (SE) \footnote{The functional form for the SE kernel is $K(x_i,x_j) = \sigma_f^2 \exp{\left[{(x_i-x_j)^2}/{2l^2}\right]}$, where $\{\sigma_f, \, l\}$ are now free parameters inferred by optimising a log-marginal likelihood. } kernel. Note that here we cannot extrapolate the profile beyond the range of the available data, as GP tends to retrieve broad prior regions, with no constrain on the posterior. We intend the analysis with GP as an alternate verification for the fits based on the profile themselves as the model-independent method is agnostic with no assumptions made for the electron density profile and the mass profile. Therefore, a parametric method based on assumed empirical profiles is expected to serve better if it is in a good agreement with the GP method. 

In \Cref{fig:data-fit-GP}, we show the mass reconstructions obtained using the NFW profile in both the GR (blue) and DHOST (red) scenarios and the GP (green) reconstructed result. For the GP method we firstly find a lowering of the mass profile in the inner regions, which is an immediate consequence of not excluding the inner most 3 data-points of $\Psz$ data (see \cite{Ettori:2018tus, Ghirardini:2018byi}). However, being a non-parametric estimation and that it only follows the data points, we find the GP to be in better agreement towards the outskirts of the mass profile. In the case of the A644, a non-NFW cluster, we find that the model-independent reconstruction is in better agreement with the GR case than with the DHOST scenario, which is a clear indication of our results for the latter being biased when utilising the NFW mass profile. The $\R = 1.15 \pm 0.06\, [{\rm Mpc}]$ and $\M = 4.57^{+0.73}_{-0.65}\, [10^{14}\, M_{\odot} ]$ estimated using GP here are in excellent agreement with our constraints in the GR case. In the case of A1795 however, the deviation between the GR and the DHOST cases is minimal and is also reflected in the GP reconstruction. Once again the GP based $\R = 1.12 \pm 0.04\, [{\rm Mpc}] $ and $\M = 4.24^{+0.045}_{-0.040}\, [10^{14}\, M_{\odot} ]$ are in excellent agreement with the parametric method assuming the NFW profile. 
%  (see Table 1 of \cite{Haridasu:2021hzq})

\subsection{$\Psz$ only analysis}
ndeed, one of the major advantages in our analysis is that we perform the joint-fit to the $\Tx$( or $\Px$) and the $\Psz$ data, which provides a better estimation of the mass profile to the outskirts ($R \sim 1.2 $ Mpc) of the clusters. In the screening mechanism, the GR is recovered in the innermost region of the galaxy cluster, while the gravity gradually deviates from GR in the outskirt of the cluster. Given that, the inclusion of the $\Psz$ data actually provides a very suitable opportunity to asses the mass profile in a larger radial range and hence the modifications to gravity. 

To contrast our constraints against the combined analysis of $\Psz+\Px$, we also perform an MCMC analysis using only the $\Psz$ data (without excluding the 3 inner most data points). We also show the inferred mass profiles (orange) of the same in \Cref{fig:data-fit-GP}. One can immediately notice that for the cluster A644, all the 4 contrasted scenarios are almost discordant at the face-value. This in turn indicates that a more careful assessment has to be made when combining the data sets and/or trying to assess the modifications to physics using this cluster. On the other hand, the A1795 cluster shows that all the approaches are are in very good agreement, especially in the outskirts of the cluster. However, comparing the $\Psz$ parametric fit and the GP based reconstruction we find that the former tends to show higher masses, also in agreement with the $\Px+\Psz$ analysis. The results for the posterior distributions on $\Xi_1$  are summarised in Appendix C. of \cite{Haridasu:2021hzq}.

\begin{figure}[h]
    \centering
    \includegraphics[scale=0.32]{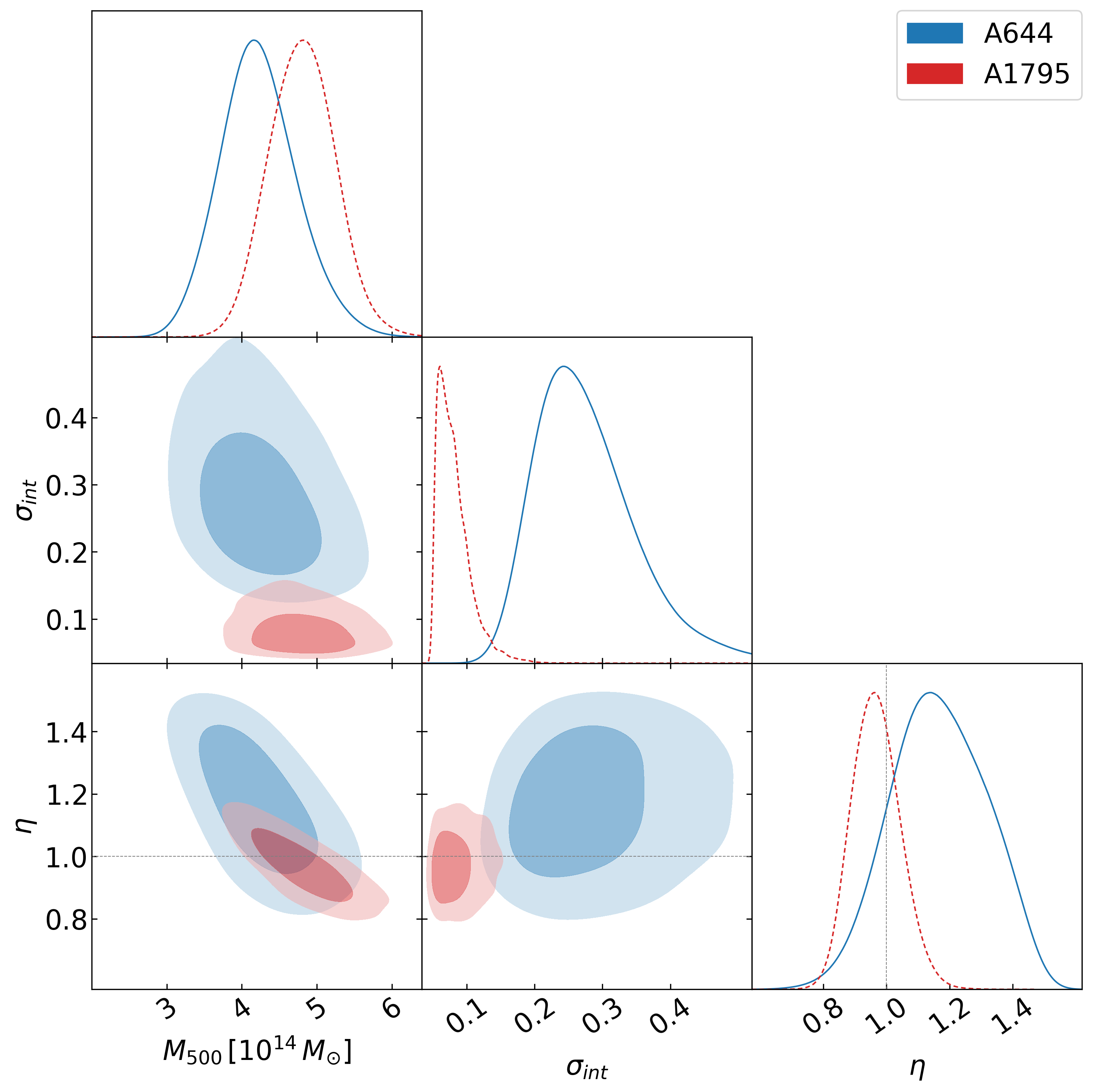}
    \caption{We compare here the posteriors for the intrinsic dispersion in the X-ray data and the scaling parameter. The dashed grey line marks the case of $\eta =1$, which implies good agreement between the $\Px$ and $\Psz$ data.}
    \label{fig:sig_eta}
\end{figure}

Within the joint analysis however, one needs to check internally for a consistency between the X-ray and SZ data, see Appendix A of \cite{Ghirardini:2018byi} for more details. Therefore in a similar approach we introduce an intrinsic dispersion ($\sigma_{\rm P, int}$) as a free parameter which is scaled by the modelled pressure ($\sigma_{P, \rm int} \sim P^{\rm model} \sigma_{\rm int}$) (Appendix D of \cite{Ghirardini:2017apw}), added in quadrature to the error and inferred through the MCMC analysis. Alongside the dispersion, we also include a rescaling parameter $\eta$, such that $P^{\rm data}_{\rm X} \to \eta \times P^{\rm data}_{\rm X} $. We show the posterior distributions of these two consistency parameters in \Cref{fig:sig_eta} for the A644 and A1795 clusters. Clearly, both clusters show a good agreement between $\Psz$ and $\Px$ data, while A644 shows a larger dispersion within the $\Px$ data points. This in turn provides additional correlations with $\M$ and increases the error-bars, however note also that we fix $\eta =1$ in the main analysis. Please refer to original analysis in \cite{Ettori:2018tus, Ghirardini:2017apw, Ghirardini:2018byi}, where these arguments are elaborated.

In summary, as we have shown here, having the galaxy cluster physics well-mapped in a larger radial range and having several of these objects also distributed along the redshift will be crucial to assess the constraints on the $\Xi_1$ in individual clusters and its time evolution. We intend to explore the same with soon to be available future data. For this goal, NIKA2 SZ observations could be helpful to extend this analysis. 

\section*{Acknowledgements}
Authors are grateful to the `mm Universe $@$NIKA2' conference for providing an oppurtunity to present the work. Authors acknowledge receiving useful comments and feedback on the use of data from Stefano Ettori and Dominique Eckert. BSH is supported by the INFN INDARK grant. PK, MDP, VC,RM acknowledge support from
INFN/Euclid Sezione di Roma. PK, MDP and RM also acknowledge support from Sapienza Universit\'a di Roma thanks to Progetti di Ricerca Medi 2018, prot. RM118164365E40D9 and 2019, prot. RM11916B7540DD8D. 
% % We acknowledge the use of publicly available python packages: \texttt{numpy, scipy, \& cmath}.
% \SH{should we remove acknowledgements? }

%
% BibTeX or Biber users please use (the style is already called in the class, ensure that the "woc.bst" style is in your local directory)
% \bibliography{name or your bibliography database}

% \newpage
% \bibliographystyle{woc.bst}
\bibliography{bibliography}

\begin{thebibliography}{17}

\bibitem{Haridasu:2021hzq}
B.S. Haridasu, P.~Karmakar, M.~De~Petris, V.F. Cardone, R.~Maoli (2021),
  \texttt{2111.01101}

\bibitem{Crisostomi:2016czh}
M.~Crisostomi, K.~Koyama, G.~Tasinato, JCAP \textbf{1604}, 044 (2016),
  \texttt{1602.03119}

\bibitem{Achour:2016rkg}
J.~Ben~Achour, D.~Langlois, K.~Noui, Phys. Rev. \textbf{D93}, 124005 (2016),
  \texttt{1602.08398}

\bibitem{Babichev:2013usa}
E.~Babichev, C.~Deffayet, Class. Quant. Grav. \textbf{30}, 184001 (2013),
  \texttt{1304.7240}

\bibitem{Navarro:1995iw}
J.F. Navarro, C.S. Frenk, S.D.M. White, Astrophys. J. \textbf{462}, 563 (1996),
  \texttt{astro-ph/9508025}

\bibitem{Vikhlinin:2005mp}
A.~Vikhlinin, A.~Kravtsov, W.~Forman, C.~Jones, M.~Markevitch, S.S. Murray,
  L.~Van~Speybroeck, Astrophys. J. \textbf{640}, 691 (2006),
  \texttt{astro-ph/0507092}

\bibitem{Sakstein:2016ggl}
J.~Sakstein, H.~Wilcox, D.~Bacon, K.~Koyama, R.C. Nichol, JCAP \textbf{1607},
  019 (2016), \texttt{1603.06368}

\bibitem{Wilcox:2015kna}
H.~Wilcox et~al., Mon. Not. Roy. Astron. Soc. \textbf{452}, 1171 (2015),
  \texttt{1504.03937}

\bibitem{Terukina:2013eqa}
A.~Terukina, L.~Lombriser, K.~Yamamoto, D.~Bacon, K.~Koyama, R.C. Nichol, JCAP
  \textbf{04}, 013 (2014), \texttt{1312.5083}

\bibitem{Ettori:2018tus}
S.~Ettori, V.~Ghirardini, D.~Eckert, E.~Pointecouteau, F.~Gastaldello,
  M.~Sereno, M.~Gaspari, S.~Ghizzardi, M.~Roncarelli, M.~Rossetti, Astron.
  Astrophys. \textbf{621}, A39 (2019), \texttt{1805.00035}

\bibitem{Saito:2015fza}
R.~Saito, D.~Yamauchi, S.~Mizuno, J.~Gleyzes, D.~Langlois, JCAP \textbf{1506},
  008 (2015), \texttt{1503.01448}

\bibitem{Sakstein:2015aac}
J.~Sakstein, Phys. Rev. \textbf{D92}, 124045 (2015), \texttt{1511.01685}

\bibitem{Sakstein:2015zoa}
J.~Sakstein, Phys. Rev. Lett. \textbf{115}, 201101 (2015), \texttt{1510.05964}

\bibitem{Saltas:2018mxc}
I.D. Saltas, I.~Sawicki, I.~Lopes, JCAP \textbf{05}, 028 (2018),
  \texttt{1803.00541}

\bibitem{Haridasu:2018gqm}
B.S. Haridasu, V.V. Lukovi\'c, M.~Moresco, N.~Vittorio, JCAP \textbf{10}, 015
  (2018), \texttt{1805.03595}

\bibitem{Ghirardini:2018byi}
V.~Ghirardini et~al., Astron. Astrophys. \textbf{621}, A41 (2019),
  \texttt{1805.00042}

\bibitem{Ghirardini:2017apw}
V.~Ghirardini, S.~Ettori, D.~Eckert, S.~Molendi, F.~Gastaldello,
  E.~Pointecouteau, G.~Hurier, H.~Bourdin, Astron. Astrophys. \textbf{614}, A7
  (2018), \texttt{1708.02954}

\end{thebibliography}

\end{document}